\documentclass[12pt]{article}
\usepackage{graphicx}
\usepackage{amssymb}
\usepackage{amsmath}
\usepackage{hyperref}
\usepackage[dvipsnames]{xcolor}
\usepackage[a4paper,margin=1in]{geometry}
\usepackage{cite}
\usepackage{parskip}
\usepackage{titlesec}
\usepackage{tikz}
\usepackage{tfrupee}
\usepackage{orcidlink}
\usepackage{authblk}
\usetikzlibrary{shapes.geometric, arrows.meta, positioning}

\title{\color{blue}{Cost-effective designs for Next-Generation Radio Telescopes}}
\author[1]{G B Raghavkrishna\,\orcidlink{0009-0002-8647-6861}}
\author[2]{Deeptangshu Banik}
\author[2]{Dr. B. Ravi Kumar\,\orcidlink{0000-0002-2723-8428}}
\author[2]{Mr. D. Veeraswamy\,\orcidlink{0009-0003-6447-9698}}
\affil[1]{Department of Physics, Indian Institute of Technology Madras, Chennai, India}
\affil[2]{Department of Electronics and Communication Engineering, Institute of Aeronautical Engineering, Hyderabad, India}

\affil[ ]{\textbf{E-mails :\hspace{1mm}}\texttt{ph24c033@smail.iitm.ac.in}, \texttt{22951a0448@iare.ac.in}, \texttt{b.ravikumar@iare.ac.in}, \texttt{d.veeraswamy@iare.ac.in}}

\date{}

\begin{document}
\maketitle
\section{\underline{Abstract}}
Radio astronomy has entered its golden era, with many revolutionary facilities such as SKA, ngVLA, and LOFAR2.0 coming online in the next decade. These facilities are certain to redefine radio astronomy. However, on smaller scales—such as at institutional or amateur levels—radio astronomy is still mostly practiced for educational or personal interest. The primary reason small-scale radio astronomical experiments rarely produce cutting-edge scientific results is the limitation of funding available for procuring or developing components similar to those used in professional-grade facilities. A second major reason is the lack of tools and skills required for simulating components for a complete telescope. In this work, we address the first of these challenges and suggest novel ideas and designs for cost-effective, next-generation radio telescopes that can be built at small scales. We also describe the observational strategies required to produce cutting-edge scientific results with these telescopes—results comparable to those from professional facilities.\\
\textbf{\underline{Keywords\hspace{1mm}}:} \textbf{Antenna Design}; \textbf{Circuit Design}; \textbf{Microstrip Lines}; \textbf{Radio Astronomy}; \textbf{Radio Telescope};

\section{\underline{Introduction}}
Ever since Arno Penzias and R. Wilson started using antennas to peer beyond the atmosphere, we can essentially say that radio astronomy was born [$\color{blue}{1}$]. In scenarios where other frequency bands of the electromagnetic spectrum such as gamma-ray, X-ray, ultraviolet, optical, and infrared fail, the radio band prevails. Electromagnetic waves in the ultraviolet, optical, and infrared frequencies are obstructed by the Earth’s atmosphere, especially the troposphere, while gamma-ray and X-ray frequencies are blocked by the atmosphere. By contrast, the radio band is the only part of the electromagnetic spectrum that allows us to observe from the surface of the Earth over a wide range of frequencies, except for the ionosphere blocking radio waves below 10 MHz and at very high frequencies where atmospheric molecules start absorbing the incoming radiation from space.

The single most important instrument in the field of radio astronomy is the antenna, and many variants have been used, with the most widely used being dipole, Yagi-Uda, and parabolic dishes. A radio telescope is essentially a receiver system consisting of an antenna, amplifier, filter, digitizer, and storage systems. The simplest radio telescope design is presented as a block diagram below:
\begin{figure}[h!]
    \centering
    \includegraphics[width=1.0\linewidth]{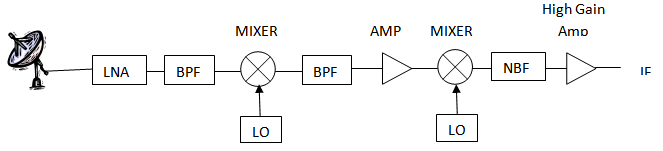}
    \caption{Block diagram of a simple super heterodyne receiver system used in a radio telescope [$\color{blue}{2}$]}
\end{figure}\\
Radio telescopes are of two types: single-dish telescopes, which have a parabolic or circular dish with very large diameters such as the Parkes 64m Radio Telescope [$\color{blue}{3}$], Effelsberg 100m Radio Telescope [$\color{blue}{4}$], FAST 500m Radio Telescope [$\color{blue}{5}$], etc.; and interferometers, in which the signals from different telescopes are superimposed so they all act as a single large telescope. Examples include the GMRT [$\color{blue}{6}$], VLA [$\color{blue}{7}$], LOFAR [$\color{blue}{8}$], ASKAP [$\color{blue}{9}$], etc.

The lowest value of intensity or flux density of electromagnetic radiation that can be observed by a detector is called its sensitivity, and in the case of a single-dish telescope it is represented as [$\color{blue}{10}$]:
\begin{gather*}
    \boxed{\sigma = \dfrac{A}{\sqrt{t_{Obs}\Delta\nu}}}\longrightarrow\color{blue}{(1)}
\end{gather*}
Here, $A$ is a system temperature-dependent proportionality constant, $t_{Obs}$ is the integration/observation time, and $\Delta\nu$ is the bandwidth of observation. Based on this equation, we can see that we need to increase the time of observation and the bandwidth to increase the sensitivity of the telescope.

All modern telescopes face the issue of insufficient sensitivity to probe very faint signals, such as the 21 cm signal from the cosmic dawn [$\color{blue}{11-13}$] and the CMB signals in the Rayleigh-Jeans part of the CMB blackbody spectrum [$\color{blue}{14-17}$]. To detect signals from these extremely distant and faint sources, we need to achieve sensitivities on the order of $\mu$K or nK.

In this work, we describe how this can be achieved on a small scale, in a cost-effective manner, while remaining scientifically relevant.

The organization of the paper is as follows: In Section 2, we describe various antenna designs that can be used for increasing the gain in the first stage of the telescope; in Section 3, we discuss how the analog and digital backend circuits can be designed to increase throughput and sensitivity, and we present novel circuit topologies to make the telescope a multipurpose system; in Section 4, we describe important science cases that can be probed with our telescope setup; in Section 5, we conclude and offer future directions.

\section{\underline{Cost-Effective Designs For High Gain Antennas}}
Since the dawn of radio astronomy, different antenna designs have been used to observe various astronomical targets with differing degrees of sensitivity and resolution. In this section, we review existing designs and suggest cost-effective designs that can be implemented at the grassroots level, such as in amateur radio astronomy or institutional setups. Despite being cost-effective, these designs still offer high gain comparable to antennas used professionally in observatories.

Parabolic antennas have been used in major facilities such as GMRT [$\color{blue}{6}$], VLA [$\color{blue}{7}$], ATCA [$\color{blue}{18}$], ASKAP [$\color{blue}{9}$], Parkes 64-m Telescope [$\color{blue}{3}$], GBT [$\color{blue}{19}$], MeerKAT [$\color{blue}{20}$], SKA-Mid [$\color{blue}{21}$], etc. They provide the advantage of offering higher gain, good beam convergence, easy maneuverability, and ultra-wide band observational capabilities ($\sim$100 MHz in GMRT up to $\sim$1000 MHz with submillimetre telescopes such as ALMA [$\color{blue}{23}$]).

Dipole antennas have also been widely used in major facilities, especially in the low-frequency bands, such as MWA [$\color{blue}{24}$], LOFAR [$\color{blue}{8}$], SKA-Low [$\color{blue}{22}$], etc. Antenna types such as log-periodic, Yagi-Uda, and horn antennas have also been used for radio astronomical purposes, yet they are not used in large-scale facilities [$\color{blue}{25}$].

Here we suggest the following designs, which are cost-effective and have high gain and sensitivity:
\begin{enumerate}
    \item Parabolic dish antennas typically provide the highest gain and best beam shape, making them the first choice. Parabolic dishes can be made in a cost-effective manner using wire meshes, offering very high gain and allowing sensitive and resolved observations in a wide frequency band covering $\sim$50 MHz to $\sim$1500 MHz (such as GMRT). Despite these advantages, there are some disadvantages, such as the construction complexity of weaving the wires together, which is high and requires industrial-grade manufacturing facilities. Additionally, the maneuvering systems for large parabolic dishes require large stepper motors. Fully plated dishes are even more complex to build on a small scale. Thus, for institutional astronomy lab setups, wire mesh parabolic dish antennas are the best-suited option.
    \item Dipole antennas act as the best cost-effective yet sensitive option for low-frequency bands ($\sim$10 MHz to $\sim$300 MHz). They offer wideband observations, do not require large spaces, and lack moving parts, unlike parabolic dishes, which require frequent maintenance. They can be easily constructed using aluminum rods and wires, making them extremely cost-effective. Simple dipole antenna designs that can be constructed at the amateur or institutional level do not offer high resolution and sensitivity. To achieve higher performance, more complex designs are needed, which is possible at the institutional level but not at the amateur level. Also, to steer the beam of the dishes across the sky, electronic beamforming is required, which might not always be feasible even at the institutional level.
    \item Horn antennas offer the best compromise between dipole antennas and parabolic dish antennas, as they are easy to construct using aluminum foil and sheets and offer wideband observation capabilities along with high gain across the entire band. Conical horn antennas specifically offer circular beams, allowing us to observe the sky without beam smearing effects. These conical horn antennas are also easy to move and do not require powerful motors compared to parabolic dishes. Their feeds are situated below the apex of the cone, so there is no scattering of incoming radio waves due to the feed support, as occurs in parabolic dishes. They also possess very low side lobes, minimizing side lobe leakage. These features allow conical horn antennas to be used at both amateur and institutional levels, enabling professional-grade scientific research. Their only minor disadvantages are large sizes for low frequencies and the need for plano-convex lenses to achieve very high gains, which increases design and construction complexity.
    \item Log-periodic antennas, Yagi-Uda antennas, and other designs also find usage in radio astronomy, but they are mostly outdated and do not find much use in modern times, with notable exceptions such as NASA's Radio JOVE Project [$\color{blue}{25}$], aimed at detecting Jovian and other planetary radio emissions (using Yagi-Uda antennas), and the Gauribandur Pulsar System (GAPS), which uses log-periodic dipole antennas [$\color{blue}{26-29}$]. They can also be used for making cost-effective telescopes, but their applicability is limited.
\end{enumerate}
For the rest of the paper, we focus on conical horn antennas because they are the most practical choice for being cost-effective in design and construction, yet scientifically capable on par with professional large-scale observatories.

\section{\underline{Circuit Designs}}
The most important part of any telescope is its detector, and in the case of radio telescopes, the back-end receiver circuitry detects incoming radio waves falling onto the antenna and records them as time-varying voltages. The most general single-dish telescope design is given as a block diagram in Figure ($\color{blue}{1}$). Many novel circuits have been developed over the years, ranging from modern low-noise, power-efficient transistors such as HEMTs, using microstrip lines and slotlines, to integrated circuits that are cryogenically cooled [$\color{blue}{30}$].

In comparison to antennas, circuit design is much more complicated, and a wide variety of possibilities and designs are available to choose from. These reasons make the development process a bit overwhelming at the amateur or small-scale institutional level. Covering the full depth of analog and digital circuit design for the radio telescope system is a formidable task, which we do not attempt in this work. Instead, we restrict the discussion to a few methodologies that can be implemented in a cost-effective yet scientifically robust manner.

The most important aspect to consider while designing receiver circuits is to ensure that they do not add too much noise to the system. This noise contributed by the circuitry is measured in terms of the system temperature, which is governed by the Friis equation, described as follows [$\color{blue}{10}$, $\color{blue}{31}$]:
\begin{gather*}
    \boxed{T_{eq} = T_{1} + \dfrac{T_{2}}{G_{1}} + \dfrac{T_{3}}{G_{1}G_{2}} + \cdots} \longrightarrow\color{blue}{(2)}
\end{gather*}
Here, $T_{i}$ and $G_{i}$ represent the individual noise temperature and gain of the $i^{th}$ component in the receiver circuit, and in terms of the noise factor this equation is expressed as:
\begin{gather*}
    \boxed{F_{Total} = F_{1} + \dfrac{F_{2}-1}{G_{1}} + \dfrac{F_{3}-1}{G_{1}G_{2}} + \dfrac{F_{4}-1}{G_{1}G_{2}G_{3}} + \cdots + \dfrac{F_{n}-1}{G_{1}G_{2}\cdots G_{n-1}}} \longrightarrow\color{blue}{(3)}
\end{gather*}
Here, $F_{i}$ and $G_{i}$ are the noise factor and available power gain, respectively, of the $i^{th}$ stage, and $n$ is the number of stages. Both magnitudes are expressed as ratios, not in decibels.

For upcoming radio telescopes, the science targets are ambitious, yet practically achievable if we are able to minimize the noise temperature significantly. Very low temperatures are achieved by using integrated circuit technology such as Monolithic Microwave Integrated Circuits (MMICs) [$\color{blue}{32}$], which combine all the individual components into a single IC chip, thereby providing very low power consumption and very low propagation losses. Using cryogenic cooling systems based on liquid nitrogen or liquid helium allows us to achieve the lowest possible noise temperatures, and these are actively used in high-frequency telescopes such as ALMA [$\color{blue}{23}$], APEX [$\color{blue}{33}$], etc. Liquid nitrogen cryogenic cooling systems, such as circulator-based systems and cryostats, are implementable at the institutional level, although they are a bit expensive and cumbersome at the amateur level. The use of liquid helium is expensive and complicated at both amateur and institutional levels.

In this work, we emphasize the use of microstrip lines, which provide the best non-integrated circuit-based option that is cost-effective and easy to handle, yet offers comparable performance [$\color{blue}{34}$]. Microstrip lines act as waveguides that can be used to design amplifiers [$\color{blue}{35}$], oscillators [$\color{blue}{36}$], filters [$\color{blue}{37}$], mixers [$\color{blue}{38}$], correlators [$\color{blue}{39}$], power couplers [$\color{blue}{40}$], and many other components on the same substrate, with a good form factor similar to what is done in ICs on a small scale. Along with circuit designs, the circuit topologies and analysis methods also need to be developed efficiently. Here we have considered microstrip lines because they are easily available, easy to handle, and cost-effective.

Our data acquisition methodology and consequently the receiver circuit topology are based on what has been implemented in LOFAR [$\color{blue}{41}$], MeerKAT [$\color{blue}{42}$], and the Gauribidanur Radio Telescope [$\color{blue}{29}$]. In MeerKAT, the entire array is used as an interferometer, and it can simultaneously function as a collection of individual single-dish telescopes. The advantage of using these antennas in single-dish mode is that we can perform coherent combination of the data recorded by each of the antennas. From Equation ($\color{blue}{1}$), we observe that the sensitivity of the single-dish telescope is inversely proportional to the square root of the observation time. Thus, we plan to build multiple antennas with such high precision that they are essentially clones of each other. This is essential to ensure that the noise contributed by each antenna (in accordance with Equation ($\color{blue}{2}$) or ($\color{blue}{3}$)) is nearly identical, with negligible deviations. We also intend to keep them close enough that the ionosphere does not vary appreciably between the antennas, so that each observes the same ionosphere at any given instant. By coherently combining the voltages recorded by each antenna, after correcting for propagation delays between them, we can achieve very high sensitivities. For example, if we observe a particular location in the sky using the aforementioned setup, where we make observations using 20 antennas for 20 hours each, we obtain an effective integration time of 400 hours, enhancing the sensitivity of the entire system by a factor of 20.

Similarly, in LOFAR, they perform sidereal averaging, in which the visibilities recorded on several sidereal days are added and averaged if the data points corresponding to individual sidereal days are overlapping or very close to each other up to a certain threshold. They then image these visibilities, which also makes the image more sensitive.

In the case of the Gauribidanur Radio Telescope, they have developed the Gauribidanur Pulsar System (GAPS), in which they use Yagi-Uda antennas and record their individual timeseries voltage data, then synthetically add phase differences to the signals recorded by each antenna to create a digitally generated, electronically steered beam. This is much easier to perform than the analog live beamforming done in phased array interferometers such as LOFAR, MWA, etc.

We have designed a novel circuit topology that essentially allows our telescope to work in dual mode simultaneously, achieved on the circuit board itself rather than later in computer post-processing. The circuit block diagram is given as follows:
\begin{figure}[htp!]
    \centering
    \includegraphics[width=1.0\linewidth]{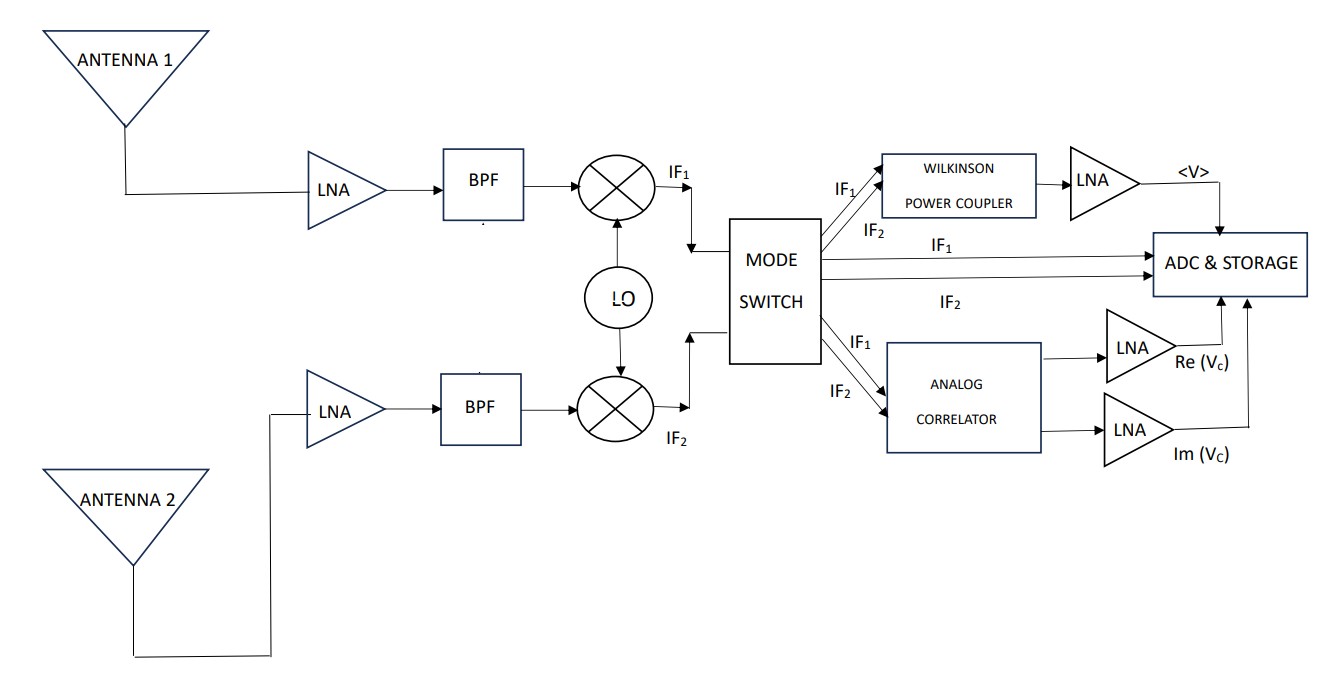}
    \caption{Block diagram of the proposed telescope, which operates in multiple modes}
\end{figure}

Here, the signals recorded by the two antennas are amplified and then filtered using a bandpass filter. A local oscillator generates the reference frequency for both signals, and they are all fed to the corresponding mixers, which convert them into the intermediate frequencies (IFs). These two IFs are then sent to the mode switch, which allows us to switch between three different modes: single-dish mode with coherent combination; interferometer; and both as single dish and interferometer simultaneously. In the case of coherent combination, the IFs are sent to the Wilkinson power coupler, which couples the IFs and produces the coherently combined final signal with the same power and no phase differences. In the case of interferometer, the IFs are sent to the analog correlator, which performs on-board correlation and provides the real and imaginary parts of the correlated visibilities. In dual mode, the IFs are simultaneously sent through both of the above-mentioned paths. All of these signals are digitized using ADCs and stored in a computer or storage device. We intend to construct all of the above-mentioned components primarily with microstrip lines and will enclose them in a Faraday cage setup made using simple metallic boxes to ensure the noise is reduced sufficiently and the components are not affected by stray fields. In the case of time-domain observations, the data recorded from the individual antennas are directly sent to ADCs and storage through the mode switch, allowing us to perform digital offline beamforming similar to GAPS.

This receiver topology allows us to achieve very low noise levels while remaining cost-effective and easy to implement at both amateur and institutional levels.

\section{\underline{Science Cases}}
In this section, we describe the possible science cases that are the prime objectives of upcoming radio telescopes and are achievable with our cost-effective telescope design.

\subsection{\underline{CMB Studies}}
The Cosmic Microwave Background has been observed over a wide range of frequencies, allowing us to probe the very beginnings of our universe. Currently, the next generation of CMB experiments, such as CMB-S4 [$\color{blue}{43}$], are underway, with some already having begun observations and others expected to join soon. Despite this, the low-frequency Rayleigh-Jeans part of the CMB blackbody spectrum is poorly observed. The TRIS experiment [$\color{blue}{14-16}$] is the only noteworthy experiment in this frequency range, with the PIXIE mission [$\color{blue}{44}$] currently being planned. In this part of the CMB spectrum, we can detect the $\mu$-type spectral distortions of the CMB spectrum [$\color{blue}{45-47}$], which occur due to energy injection into the universe in its very early stages in the redshift range of \(2\times 10^{6}>z>2\times 10^{5}\). $\mu$-type distortions are produced when Compton scattering is still efficient enough to maintain kinetic equilibrium in the plasma, meaning the energy is distributed according to a thermal distribution. This can be measured by measuring the temperature of the CMB at different frequencies and plotting them to observe noticeable deviations from the ideal Planck blackbody spectrum. In this frequency range, the $\mu$-type CMB spectral distortions are more pronounced than in other regions of the CMB spectrum.

The variation in intensity beyond the Planck blackbody spectrum due to $\mu$-type distortions is given by [$\color{blue}{45}$]:
\begin{gather*}
    \boxed{\Delta I_{\nu}^{(\mu)} = \mu\cdot M(\nu)}\longrightarrow\color{blue}{(4)}
\end{gather*}
Here, $\mu$ is the dimensionless chemical potential parameter, and $M(\nu)$ is the spectral shape function for the $\mu$-type distortion, given by (for a given CMB temperature $T$):
\begin{gather*}
    \boxed{M(\nu) = \dfrac{2h\nu^{3}}{c^{2}}\dfrac{e^{\frac{h\nu}{k_{B}T}}}{(e^{\frac{h\nu}{k_{B}T}}-1)^{2}}\left(\dfrac{\frac{h\nu}{k_{B}T}}{2.19} -1\right)}\longrightarrow\color{blue}{(5)}
\end{gather*}
Our proposed telescope idea with the conical horn antenna is best suited for this purpose, and if we observe a particular region of the sky for an integration time of 3000 hours, we can achieve $\mu$K sensitivities (in accordance with Equation ($\color{blue}{1}$)) using 93 such antennas for an observing bandwidth of 10 MHz. Thus, using our proposed telescope design, we can probe the very early universe in a cost-effective manner.

\subsection{\underline{21 cm Cosmology}}
Similar to observing the CMB, our telescope can be used for 21 cm cosmology [$\color{blue}{48}$, $\color{blue}{49}$], and we can measure the 21 cm global temperature using the same setup. The hydrogen line—also known as the 21-centimeter line or HI line—is a specific spectral feature resulting from a change in the energy state of individual, electrically neutral hydrogen atoms. This line arises due to a spin-flip transition [$\color{blue}{50}$], where the electron’s spin orientation reverses with respect to the proton’s spin. This transition occurs between the two hyperfine levels of hydrogen’s 1s ground state. The electromagnetic radiation associated with this process has a precise frequency of 1420.405751768(2) MHz (or 1.42 GHz), which corresponds to a vacuum wavelength of 21.106114054160(30) centimeters. By mapping the tomography of these emissions, we can probe the structure of the universe and its evolution during the cosmic dark ages, and it also allows us to probe the first stars and galaxies and their evolution. The relation between the redshift of the signal and the frequency of observation is given by the Doppler shift formula as [$\color{blue}{48}$]:
\begin{gather*}
    \boxed{\nu\hspace{1mm}[MHz]= \dfrac{1420.4\hspace{1mm}MHz}{1+z}}\longrightarrow\color{blue}{(6)}
\end{gather*}
Here, $z$ is the redshift of the signal. By expressing the above relation in terms of temperature, we observe the frequency dependence on the 21 cm brightness temperature, represented as follows:
\begin{figure}[htp!]
    \centering
    \includegraphics[width=0.9\linewidth]{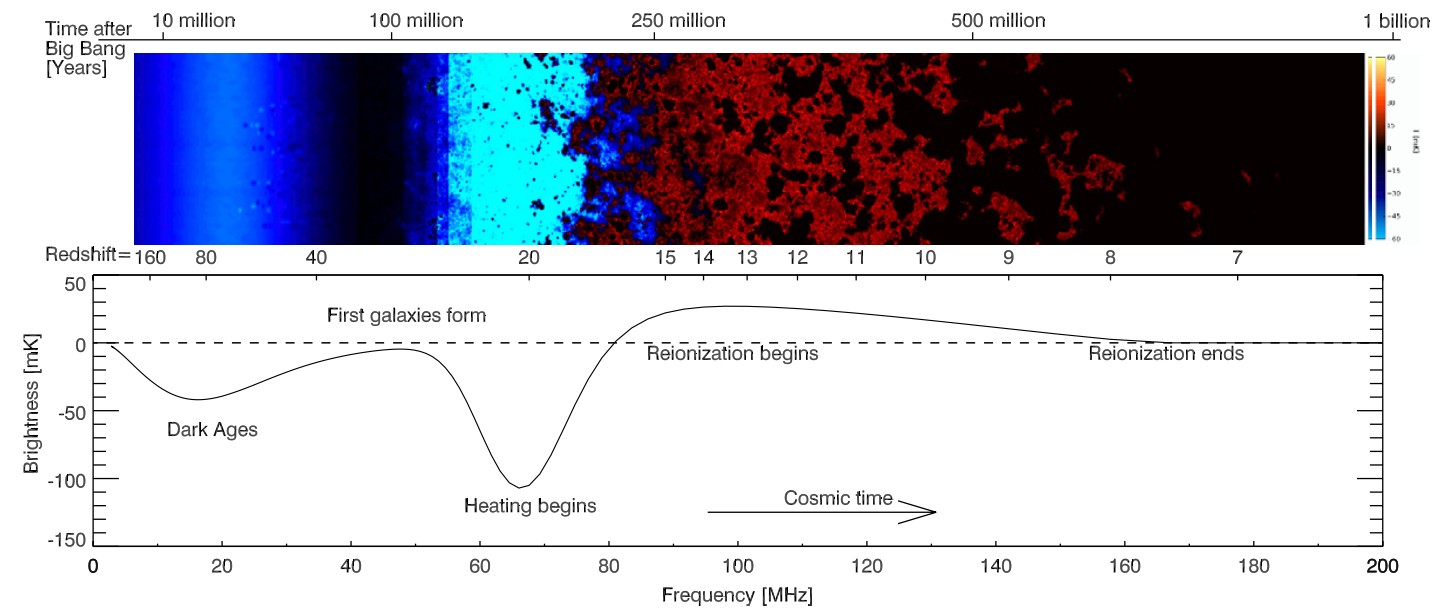}
    \caption{Variation of 21 cm global brightness temperature with frequency [$\color{blue}{48}$]}
\end{figure}

Since we are able to achieve sensitivity levels of $\mu$K in the case of the CMB, we can easily achieve mK sensitivities with the same setup, which is comparable with other existing and proposed experiments such as SARAS [$\color{blue}{51}$], EDGES [$\color{blue}{52}$], REACH [$\color{blue}{53}$], HERA [$\color{blue}{54}$], etc.

While the global temperature gives the sky-averaged temperature and its variation with frequency, we can obtain more information about the large-scale structure of the universe with the help of the 21 cm power spectrum, which is given by the 2-point correlation function for the 21 cm brightness temperature fluctuations as follows [$\color{blue}{55-57}$]:
\begin{gather*}
    \boxed{\left\langle \tilde{\delta T}_b(\mathbf{k}) \, \tilde{\delta T}_b^*(\mathbf{k}') \right\rangle = (2\pi)^3 \delta^3(\mathbf{k} - \mathbf{k}') \, P_{21}(k)}\longrightarrow\color{blue}{(7)}
\end{gather*}
Here, $\tilde{\delta T}_b(\mathbf{k})$ is the Fourier transform of the 21 cm brightness temperature fluctuation field, and $P_{21}(k)$ is the 21 cm power spectrum as a function of wavenumber $k$.

The power spectrum is estimated using the visibilities, and thus by using our telescope in dual mode, we can measure both the global temperature and the power spectrum simultaneously. Thus, our proposed telescope design offers a great advantage by providing both summary statistics at the same time with very high sensitivity, while still remaining cost-effective.

\subsection{\underline{Pulsar and Transient Studies}}
Pulsars are rapidly rotating neutron stars [$\color{blue}{58-60}$ and references therein] (in some rare cases, white dwarf stars too [$\color{blue}{61-63}$]) which emit intense radiation periodically towards Earth, so that we observe a periodic pulsed emission when plotting intensity as a function of time. To detect these pulses, we need good time resolution to resolve the pulse profile well. In this regard, our proposed telescope can observe pulsars and record the time series in all three modes, especially using the direct data storing mode and then performing digital beamforming on the stored data.

Similarly, other time-dependent signals can be detected from transient sources such as FRBs [$\color{blue}{64-66}$] using the same observation methods with our proposed telescope setup. We can employ sparse sampling [$\color{blue}{67}$, $\color{blue}{68}$] in the data while it is being stored, reducing the volume of data recorded while still observing these transient signals with good signal-to-noise ratio (SNR).

\section{\underline{Conclusions}}
In this work, we have reviewed the existing technologies and methodologies used in radio astronomy and described various cost-effective designs and methods to be used for next-generation radio telescopes, which can be set up on small scales such as amateur or institutional level and can deliver science results comparable to professional large-scale observatories. These are as follows:
\begin{enumerate}
    \item We propose to use conical horn antennas, as they combine the best features of large parabolic dishes and dipole antennas. They are also easy to design and manufacture, with costs not exceeding \rupee 2000.
    \item For analog receiver systems, we propose using microstrip lines and dielectric substrates with novel circuit designs, as they offer low noise comparable to ICs and are easy to design and fabricate. The total cost for the substrates and lines will be around \rupee 5000. For the ADCs, we can use software-defined radios (SDRs), costing around \rupee 6000. The data can be stored in HDDs, which cost around \rupee 5000 for 2TB. Thus, the total receiver circuitry cost will be around \rupee 16000 and not exceed \rupee 20000.
    \item We propose multi-mode observation strategies, allowing us to increase the sensitivities of the telescope to probe very faint emissions such as the CMB, 21 cm brightness temperature, and to use data storage and processing strategies to probe transient events.
\end{enumerate}
Hence, our proposed telescope setup will provide scientific results comparable to large-scale professional observatories on a budget not more than \rupee 25000 (for a two element telescope), making it feasible to be adopted at amateur and institutional levels. Before setting up the telescope, we can perform wideband observations using a simple dipole antenna and check which bands are free from RFI; upon doing so, this telescope can be used even in cityscapes and will offer the best results in those uncontaminated bands in large urban areas.

These designs are currently under development, and in future work, we will detail the simulations and practical implementation of this telescope, describing how it allows us to achieve significant scientific results.
\section{\underline{References}}
\begin{enumerate}
    \item Penzias, A.A.; R. W. Wilson (July 1965). "A Measurement Of Excess Antenna Temperature At 4080 Mc/s". \href{https://ui.adsabs.harvard.edu/abs/1965ApJ...142..419P/abstract}{Astrophysical Journal Letters. 142: 419–421}
    \item Daniyan, Lanre. (2011). "THE CONCEPT OF RADIO TELESCOPE RECEIVER DESIGN."\href{https://www.ripublication.com/irph/ijece/ijecev4n4__10.pdf}{International Journal of Electronics and Communication Engineering. ISSN 0974-2166 Volume 4, Number 4 (2011), pp. 453-460} 
    \item For more details about Parkes 64m Telescope check \href{https://www.csiro.au/en/about/facilities-collections/atnf/parkes-radio-telescope-murriyang}{$\color{blue}{here}$}.
    \item For more details about Effelsberg 100m Radio Telescope check \href{https://www.mpifr-bonn.mpg.de/en/effelsberg}{$\color{blue}{here}$}.
    \item For more details about FAST 500m Radio Telescope check \href{https://fast.bao.ac.cn/}{$\color{blue}{here}$}.
    \item For more details about GMRT check \href{https://www.gmrt.ncra.tifr.res.in/}{$\color{blue}{here}$}.
    \item For more details about VLA check \href{https://science.nrao.edu/facilities/vla}{$\color{blue}{here}$}.
    \item For more details about LOFAR check \href{https://science.astron.nl/telescopes/lofar/}{$\color{blue}{here}$}.
    \item For more details about ASKAP check \href{https://www.csiro.au/en/about/facilities-collections/atnf/askap-radio-telescope}{$\color{blue}{here}$}.
    \item Antenna Theory: Analysis and Design, 4th Edition, Constantine A. Balanis, \href{https://www.wiley.com/en-us/Antenna+Theory%3A+Analysis+and+Design%2C+4th+Edition-p-9781118642061}{ISBN: 978-1-118-64206-1, February 2016, 1104 pages}
    \item Bowman, Judd D.; Rogers, Alan E. E.; Monsalve, Raul A.; Mozdzen, Thomas J.; Mahesh, Nivedita (1 March 2018). "An absorption profile centred at 78 megahertz in the sky-averaged spectrum". \href{https://www.nature.com/articles/nature25792}{Nature. 555 (7694): 67–70}
    \item Singh, Saurabh; Nambissan T., Jishnu; Subrahmanyan, Ravi; Udaya Shankar, N.; Girish, B. S.; Raghunathan, A.; Somashekar, R.; Srivani, K. S.; Sathyanarayana Rao, Mayuri (28 February 2022). "On the detection of a cosmic dawn signal in the radio background". \href{https://www.nature.com/articles/s41550-022-01610-5}{Nature Astronomy. 6 (5): 607–617.}
    \item H. T. J. Bevins, A. Fialkov, E. de Lera Acedo, W. J. Handley, S. Singh, R. Subrahmanyan \& R. Barkana. "Astrophysical constraints from the SARAS 3 non-detection of the cosmic dawn sky-averaged 21-cm signal". \href{https://www.nature.com/articles/s41550-022-01825-6}{Nature Astronomy Volume 6, pages1473–1483 (2022)}
    \item Zannoni, M. search by orcid ; Tartari, A. search by orcid ; Gervasi, M. ; Boella, G. ; Sironi, G. ; De Lucia, A. ; Passerini, A. ; Cavaliere, F. "TRIS. I. Absolute Measurements of the Sky Brightness Temperature at 0.6, 0.82, and 2.5 GHz". \href{https://ui.adsabs.harvard.edu/abs/2008ApJ...688...12Z/abstract}{The Astrophysical Journal, Volume 688, Issue 1, pp. 12-23 (2008).}
    \item Gervasi, M. ; Zannoni, M. search by orcid ; Tartari, A. search by orcid ; Boella, G. ; Sironi, G. "TRIS. II. Search for CMB Spectral Distortions at 0.60, 0.82, and 2.5 GHz". \href{https://ui.adsabs.harvard.edu/abs/2008ApJ...688...24G/abstract}{The Astrophysical Journal, Volume 688, Issue 1, pp. 24-31 (2008).}
    \item Tartari, A. search by orcid ; Zannoni, M. search by orcid ; Gervasi, M. ; Boella, G. ; Sironi, G. "TRIS. III. The Diffuse Galactic Radio Emission at $\delta$ = + $42^{\circ}$\hspace{1mm}" \href{https://ui.adsabs.harvard.edu/abs/2008ApJ...688...32T/abstract}{The Astrophysical Journal, Volume 688, Issue 1, pp. 32-42 (2008).}
    \item Raghunathan, A., Subrahmanyan, R. "A measurement of the cosmic microwave background temperature 1280 MHz". \href{https://link.springer.com/article/10.1007/BF02702258}{ J Astrophys Astron 21, 1–17 (2000).}
    \item For more details about ATCA check \href{https://www.narrabri.atnf.csiro.au/}{$\color{blue}{here}$}
    \item For more details about GBT check \href{https://greenbankobservatory.org/}{$\color{blue}{here}$}
    \item For more details about MeerKAT check \href{https://www.sarao.ac.za/science/meerkat/about-meerkat/}{$\color{blue}{here}$}
    \item For more details about SKA Mid check \href{https://www.skao.int/en/explore/telescopes/ska-mid}{$\color{blue}{here}$}
    \item For more details about SKA Low check \href{https://www.skao.int/en/explore/telescopes/ska-low}{$\color{blue}{here}$}
    \item For more details about ALMA check \href{https://www.almaobservatory.org/en/about-alma/}{$\color{blue}{here}$}
    \item For more details about MWA check \href{https://www.mwatelescope.org/}{$\color{blue}{here}$}
    \item For more details about NASA's Radio JOVE Project check \href{https://radiojove.gsfc.nasa.gov/}{$\color{blue}{here}$}
    \item K. S. Bane, I. V. Barve, G. V. S. Gireesh, C. Kathiravan and R. Ramesh, "Gauribidanur Pulsar System" \href{https://ieeexplore.ieee.org/document/10118458/authors#authors}{2022 URSI Regional Conference on Radio Science (USRI-RCRS), Indore, India, 2022, pp. 1-3}
    \item Kshitij S. Bane, Indrajit V. Barve, Gantyada Venkata Satya Gireesh, Chidambaram Kathiravan, and Rajaram Ramesh "Initial results from multi-beam observations of pulsars and solar transient with the digital beamformer for the Gauribidanur Pulsar System" \href{https://www.spiedigitallibrary.org/journals/Journal-of-Astronomical-Telescopes-Instruments-and-Systems/volume-10/issue-1/014001/Initial-results-from-multi-beam-observations-of-pulsars-and-solar/10.1117/1.JATIS.10.1.014001.short}{Journal of Astronomical Telescopes, Instruments, and Systems 10(1), 014001 (29 December 2023).}
    \item Kshitij S. Bane, Indrajit V. Barve, G. V. S. Gireesh, C. Kathiravan,and R. Ramesh "Gauribidanur Pulsar System : A prototype software-based binary radio telescope for transient surveys ". \href{https://www.ursi.org/proceedings/procAT24/papers/0445.pdf}{4th URSI AT-RASC, Gran Canaria, 19-24 May 2024}
    \item G V S, Gireesh \& Bane, Kshitij \& Barve, Indrajit \& Chidambaram, Kathiravan \& Rajaram, Ramesh. (2024). "Observing Pulsars and Transients at Low Radio frequencies using 1-bit digitization." \href{https://www.researchgate.net/publication/387557496_Observing_Pulsars_and_Transients_at_Low_Radio_frequencies_using_1-bit_digitization}{6th URSI-RCRS 2024, Bhimtal, India}
    \item RF Circuit Design : Theory and Applications, 2nd Edition, Reinhold Ludwig, Gene Bogdanov \href{http://ece.wpi.edu/RF_Circuit_Design/index.html}{Prentice Hall, 2009, ISBN : 0131471376, 9780131471375}
    \item Fundamentals of Radio Astronomy : Observational Methods, 1st Edition, Jonathan M. Marr, Ronald L. Snell, Stanley E. Kurtz \href{https://www.routledge.com/Fundamentals-of-Radio-Astronomy-Observational-Methods/Marr-Snell-Kurtz/p/book/9780367575236?srsltid=AfmBOor6OKwRaKy7wcWvxhQL8VosZF3QiH1cxDxzeKnZutimiD9JNwn1}{CRC Press, 2020, ISBN : 9780367575236}
    \item Microwave Active Circuit Analysis and Design, 1st Edition, Clive Poole, Izzat Darwazeh \href{https://www.sciencedirect.com/book/9780124078239/microwave-active-circuit-analysis-and-design}{Academic Press, 2015, ISBN : 978-0-12-407823-9}
    \item For more details about APEX check \href{https://www.apex-telescope.org/ns/}{$\color{blue}{here}$}
    \item Microstrip Lines and Slotlines, Third Edition, Inder Bahl; Maurizio Bozzi; Ramesh Garg, \href{https://ieeexplore.ieee.org/document/9101138/authors#authors}{Artech, 2013, ISBN : 9781608075362}
    \item B Maruddani, M Ma'sum, E Sandi, Y Taryana, T Daniati and W Dara, "Design of two stage low noise amplifier at 2.4 - 2.5 GHz frequency using microstrip line matching network method" \href{https://iopscience.iop.org/article/10.1088/1742-6596/1402/4/044031}{Journal of Physics: Conference Series, Volume 1402, Issue 4}
    \item S.H. Ibrahim; E.-S.A. El-Badawy; H.A. El-Motanfy, "Computer-aided design of 2.6 GHz microstrip oscillator" \href{https://ieeexplore.ieee.org/document/825591/authors#authors}{Proceedings of the Tenth International Conference on Microelectronics (Cat. No.98EX186), IEEE, 06 August 2002}
    \item Orhan Yeşilyurt; Mustafa Köksal; Taha İmeci, "Microstrip band pass filter design" \href{https://ieeexplore.ieee.org/document/7129838/authors#authors}{Conference Proceedings : 2015 23nd Signal Processing and Communications Applications Conference (SIU), IEEE, May 2015}
    \item Citron, N.; Holdengreber, E.; Sorkin, O.; Schacham, S.E.; Farber, E. "High-Performance RF Balanced Microstrip Mixer Configuration for Cryogenic and Room Temperatures". \href{https://www.mdpi.com/2079-9292/11/1/102}{Electronics 2022, 11, 102.}
    \item Norhudah Seman; Marek E. Bialkowski; Siti Z. Ibrahim; Aslina Abu Bakar "Design of an integrated correlator for application in ultra wideband six-port transceivers" \href{https://ieeexplore.ieee.org/document/5172332}{Conference : 2009 IEEE Antennas and Propagation Society International Symposium, IEEE, 24 July 2009}
    \item Wang, X., Ohira, M. and Ma, Z. (2016), "Coupled microstrip line Wilkinson power divider with open-stubs for compensation". \href{https://ietresearch.onlinelibrary.wiley.com/doi/epdf/10.1049/el.2016.1819}{Electron. Lett., 52: 1314-1316}
    \item J. M. G. H. J. de Jong, R. J. van Weeren, T. J. Dijkema, J. B. R. Oonk, H. J. A. Röttgering and F. Sweijen, "Unlocking ultra-deep wide-field imaging with sidereal visibility averaging" \href{https://www.aanda.org/articles/aa/full_html/2025/02/aa52492-24/aa52492-24.html}{A98, A\&A Volume 694, February 2025} \& references therein
    \item For more details about MeerKAT Single Dish Mode check \href{https://indico.ph.tum.de/event/7160/attachments/5002/6437/Steve_Cunnington.pdf}{$\color{blue}{here}$}
    \item For more details about CMB-S4 check \href{https://cmb-s4.org/}{$\color{blue}{here}$}
    \item Alan Kogut, Eric Switzer, Dale Fixsen, Nabila Aghanim, Jens Chluba, Dave Chuss, Jacques Delabrouille, Cora Dvorkin, Brandon Hensley, Colin Hill, Bruno Maffei, Anthony Pullen, Aditya Rotti, Alina Sabyr, Leander Thiele, Ed Wollack, Ioana Zelko., "The Primordial Inflation Explorer (PIXIE) : Mission Design and Science Goals" \href{https://arxiv.org/abs/2405.20403}{Journal of Cosmology and Astroparticle Physics, Volume 2025, Issue 04, id.020, 45 pp}
    \item Hiroyuki Tashiro., "CMB spectral distortions and energy release in the early universe" \href{https://academic.oup.com/ptep/article/2014/6/06B107/1560917?login=false}{Progress of Theoretical and Experimental Physics, Volume 2014, Issue 6, June 2014, 06B107}
    \item F. Bianchini, G. Fabbian "CMB spectral distortions revisited : A new take on $\mu$ distortions and primordial non-Gaussianities from FIRAS data". \href{https://journals.aps.org/prd/abstract/10.1103/PhysRevD.106.063527}{Phys. Rev. D 106, 063527, 2022}
    \item Bryce Cyr., "CMB Spectral Distortions : A Multimessenger Probe of the Primordial Universe", \href{https://arxiv.org/abs/2406.12985v1}{eprint arXiv:2406.12985, June 2024}
    \item Jonathan R Pritchard and Abraham Loeb, "21 cm cosmology in the 21st century" \href{https://iopscience.iop.org/article/10.1088/0034-4885/75/8/086901}{2012 Rep. Prog. Phys. 75 086901} \& references therein.
    \item Rennan Barkana, Abraham Loeb, "In the beginning: the first sources of light and the reionization of the universe", \href{https://www.sciencedirect.com/science/article/abs/pii/S0370157301000199?via%3Dihub}{Physics Reports, Volume 349, Issue 2, 2001, Pages 125-238, ISSN 0370-1573}
    \item H. Hellwig, R. F. C. Vessot, M. W. Levine, P. W. Zitzewitz, D. W. Allan and D. J. Glaze, "Measurement of the Unperturbed Hydrogen Hyperfine Transition Frequency," \href{https://ieeexplore.ieee.org/document/4313902}{in IEEE Transactions on Instrumentation and Measurement, vol. 19, no. 4, pp. 200-209, Nov. 1970, doi: 10.1109/TIM.1970.4313902}.
    \item Patra, N., Subrahmanyan, R., Raghunathan, A., Udaya Shankar, N., "SARAS : a precision system for measurement of the cosmic radio background and signatures from the epoch of reionization" \href{https://link.springer.com/article/10.1007/s10686-013-9336-3}{Exp Astron 36, 319–370 (2013)}.
    \item For more details about the EDGES experiment check \href{https://www.haystack.mit.edu/astronomy/astronomy-projects/edges-experiment-to-detect-the-global-eor-signature/}{$\color{blue}{here}$}
    \item de Lera Acedo, E., de Villiers, D.I.L., Razavi-Ghods, N. et al. "The REACH radiometer for detecting the 21-cm hydrogen signal from redshift z $\approx$ 7.5–28". \href{https://www.nature.com/articles/s41550-022-01709-9}{Nat Astron 6, 984–998 (2022).}
    \item David R. DeBoer et al., "Hydrogen Epoch of Reionization Array (HERA)" \href{https://iopscience.iop.org/article/10.1088/1538-3873/129/974/045001}{PASP 129 045001, 2017}
    \item Ivelin Georgiev, Garrelt Mellema, Sambit K Giri, Rajesh Mondal, "The large-scale 21-cm power spectrum from reionization" , \href{https://academic.oup.com/mnras/article/513/4/5109/6581343?login=false}{Monthly Notices of the Royal Astronomical Society, Volume 513, Issue 4, July 2022, Pages 5109–5124}
    \item S.  Munshi, F. G.  Mertens, L. V. E.  Koopmans, A. R.  Offringa, B.  Semelin, D.  Aubert, R.  Barkana, A.  Bracco, S. A.  Brackenhoff, B.  Cecconi, E.  Ceccotti, S.  Corbel, A.  Fialkov, B. K.  Gehlot, R.  Ghara, J. N.  Girard, J. M.  Grießmeier, C.  Höfer, I.  Hothi, R.  Mériot, M.  Mevius, P.  Ocvirk, A. K.  Shaw, G.  Theureau, S.  Yatawatta, P.  Zarka, S.  Zaroubi, "First upper limits on the 21 cm signal power spectrum from cosmic dawn from one night of observations with NenuFAR" \href{https://www.aanda.org/articles/aa/abs/2024/01/aa48329-23/aa48329-23.html}{A\&A 681 A62 (2024)}
    \item Rennan Barkana and Abraham Loeb, "A Method for Separating the Physics from the Astrophysics of High-Redshift 21 Centimeter Fluctuations" \href{https://iopscience.iop.org/article/10.1086/430599}{ApJ 624 L65, 2005}
    \item A. Philippov and M. Kramer, "Pulsar Magnetospheres and Their Radiation" \href{https://www.annualreviews.org/content/journals/10.1146/annurev-astro-052920-112338#abstract_content}{Vol. 60:495-558, Annual Reviews of Astronomy \& Astrophysics, 2022}
    \item Beskin, V.S., Chernov, S.V., Gwinn, C.R. et al. "Radio Pulsars". \href{https://link.springer.com/article/10.1007/s11214-015-0173-8}{Space Sci Rev 191, 207–237 (2015)}.
    \item Zhou, S.; Gügercinoğlu, E.; Yuan, J.; Ge, M.; Yu, C. "Pulsar Glitches : A Review". \href{https://www.mdpi.com/2218-1997/8/12/641}{Universe 2022, 8, 641}.
    \item Marsh, T. R., Gänsicke, B. T., Hümmerich, S., et al.  "A radio-pulsing white dwarf binary star." \href{https://www.nature.com/articles/nature18620}{Nature, 537, 374–377 (2016)}.
    \item Buckley, D., Meintjes, P., Potter, S. et al. "Polarimetric evidence of a white dwarf pulsar in the binary system AR Scorpii". \href{https://www.nature.com/articles/s41550-016-0029}{Nat Astron 1, 0029 (2017).}
    \item Pelisoli, I., Marsh, T.R., Buckley, D.A.H. et al. "A 5.3-min-period pulsing white dwarf in a binary detected from radio to X-rays". \href{https://www.nature.com/articles/s41550-023-01995-x}{Nat Astron 7, 931–942 (2023)}.
    \item Bing Zhang, "The physics of fast radio bursts", \href{https://journals.aps.org/rmp/abstract/10.1103/RevModPhys.95.035005}{Rev. Mod. Phys. 95, 035005, 2023}.
    \item Cherry Ng, "A brief review on Fast Radio Bursts", \href{https://arxiv.org/abs/2311.01899}{arXiv:2311.01899 [astro-ph.HE]}
    \item Petroff, E., Hessels, J.W.T. \& Lorimer, D.R. "Fast radio bursts" \href{https://link.springer.com/article/10.1007/s00159-019-0116-6}{Astron Astrophys Rev 27, 4 (2019)}.
    \item Hao Shan, Jianping Yuan, Na Wang, and Zhen Wang, "Compressed Sensing Based RFI Mitigation and Restoration for Pulsar Signals", \href{https://iopscience.iop.org/article/10.3847/1538-4357/ac8003}{ApJ 935 117, 2022}
    \item Hang Yu, Luping Xu, Dongzhu Feng, Xiaochuan He, Xiaoling Ye, "A sparse representation-based optimization algorithm for measuring the time delay of pulsar integrated pulse profile" \href{https://www.sciencedirect.com/science/article/abs/pii/S1270963815001935}{Aerospace Science and Technology, Volume 46, 2015, Pages 94-103, ISSN 1270-9638}
\end{enumerate}
\end{document}